\documentstyle[12pt]{article}

\begin{document}
\begin{center}
{\Large \bf Improved scenario of baryogenesis }
\bigskip

{\large D.L.~Khokhlov}
\smallskip

{\it Sumy State University, R.-Korsakov St. 2, \\
Sumy 244007, Ukraine\\
E-mail: khokhlov@cafe.sumy.ua}
\end{center}

\begin{abstract}
It is assumed that, in the primordial plasma,
at the temperatures above the mass of electron,
fermions are in the neutral state being the superposition of
particle and antiparticle.
There exists neutral proton-electron symmetry.
Proton-electron equilibrium
is defined by the proton-electron mass difference.
At the temperature equal to the mass of electron,
pairs of neutral electrons annihilate into photons, and
pairs of neutral protons and electrons survive as
protons and electrons.

\end{abstract}

Standard scenario of baryogenesis occurs as follows~\cite{Kolb}.
In the thermodynamically equilibrium primordial plasma,
all the particles with $m\leq T$ exist in equal abundances
per spin degree of freedom. This means that, at $T>m_{p}$,
there exist approximately equal abundances of protons $p$ and
antiprotons $\bar p$. Excess of baryonic charge arises in the
processes of X, Y-bosons decays under the following conditions:
non-conservation of the baryonic charge and CP-violation.
At $T<m_{p}$, $p\bar p$ annihilate, with the excess of $p$
survives.

Let us consider another scenario of baryogenesis.
Let us assume that, in the primordial plasma,
at $T>m_e$, fermions are in the neutral state being
the superposition of particle and antiparticle
\begin{equation}
|\psi^0>=\frac{1}{\sqrt 2}(\bar \psi+\psi).
\label{eq:psi}
\end{equation}
Neutral electrons
are in the state being the superposition of electron and
positron
\begin{equation}
|e^0>=\frac{1}{\sqrt 2}(e^-+e^+).
\label{eq:en}
\end{equation}
Neutral protons
are in the state being the superposition of proton and
antiproton
\begin{equation}
|p^0>=\frac{1}{\sqrt 2}(\bar p + p).
\label{eq:pn}
\end{equation}

Let us assume that, at $T>m_e$,
there exists neutral proton-electron symmetry.
For the neutral electron, the sequence of spin transformations
$\uparrow\downarrow\uparrow\downarrow\uparrow$
transits the particle into itself.
Let us identify
the cluster of neutral electrons
$\uparrow\downarrow\uparrow\downarrow\uparrow$
with the neutral proton.
The probability of finding such a cluster is given by
\begin{equation}
w=\left(\frac{1}{2}\right)^{5}.
\label{eq:w}
\end{equation}
Proton-electron equilibrium
is defined by the proton-electron mass difference.
The proton-electron ratio is given by
\begin{equation}
\frac{N_{p}}{N_{e}}=\left(\frac{1}{2}\right)^{5}
\left(\frac{m_{e}}{m_{p}}\right)^{2}
\label{eq:NpNe}
\end{equation}

At $T=m_{e}$, pairs of neutral electrons annihilate into photons
\begin{equation}
e^0 e^0\rightarrow \gamma\gamma.
\label{eq:eg}
\end{equation}
Pairs of neutral protons and electrons survive as
protons and electrons
\begin{equation}
p^0 e^0\rightarrow p e^-.
\label{eq:pe}
\end{equation}
Birth of antiprotons and positrons
is forbidden, since this leads to $\bar pp$ and $e^+e^-$
annihilation that is inconsistent with $p e^-$ survival.
Unlike the symmetry between neutral proton $p^0$  and
neutral electron $e^0$,
there exists no symmetry between proton $p$ and
positron $e^+$ that forbids the decay of proton.

Baryon-photon ratio at $T=m_{e}$ is given by
\begin{equation}
\frac{N_{b}}{N_{\gamma}}=\frac{3}{4}\frac{N_{p}}{N_{e}}
\label{eq:NbNg}
\end{equation}
where fraction $3/4$ takes into account relation between
fermions and bosons. Calculations yield the value
$N_{b}/N_{\gamma}=6.96\times 10^{-9}$.
The observed value of $N_{b}/N_{\gamma}$ lies in the range
$2-15\times 10^{-10}$~\cite{ISSI}.
Let us assume that the most fraction of baryonic matter
decays into non-baryonic matter during the evolution of the universe.
Estimate baryon number density at $T=m_{e}$ from the modern total
mass density of the universe. According to the model of the
universe with the linear evolution law~\cite{Kh},
mass density of the universe is given by
\begin{equation}
\rho={3\over{4\pi G t^2}}.
\label{eq:rho}
\end{equation}
Modern age of the universe is given by
\begin{equation}
t_{0}=t_{Pl}\alpha\left(\frac
{T_{Pl}}{T_{0}}\right)^2.
\label{eq:age}
\end{equation}
From this the modern age of the universe is equal to $t_{0}=
1.06\times 10^{18} \ {\rm s}$,
and the modern mass density of the universe is equal to
$\rho_{0}=3.19\times 10^{-30}\ {\rm g\ cm^{-3}}$.
This value corresponds to the relativistic matter. To transit to
the usual matter it is necessary to multiply the value by a
factor of 2.
Then the baryon number density at $T=m_{e}$ is
$n_{b}=2\rho_{0}/m_{p}=3.8\times 10^{-6}\ {\rm cm^{-3}}$.
While adopting the observed photon number density as
$n_{\gamma}=550 \ {\rm cm^{-3}}$~\cite{Dolg},
the baryon-photon ratio at $T=m_{e}$ is equal to
$N_{b}/N_{\gamma}=6.92\times 10^{-9}$.

It should be noted that, in comparison with
separate fermions and antifermions,
neutral fermions in the state given by eq. (\ref{eq:psi})
have energies greater by a factor of 2.
From this cross-sections of reactions
in the primordial plasma increase.

\section*{Acknowledgements}

I thank G. Steigman for helpful comments.

\end{document}